\documentclass[aps,pre,reprint,superscriptaddress]{revtex4-2}
\usepackage{amsmath}
\usepackage{amsfonts}
\usepackage{amssymb}
\usepackage{mathrsfs}
\usepackage{mathtools}
\usepackage{xcolor}

\newcommand{\dd}{\mathrm{d}}
\newcommand{\normII}[1]{\left\|#1\right\|}
\newcommand{\normp}[1]{\left\|#1\right\|_p}
\newcommand{\normInt}[1]{\lvert#1\rvert}
\newcommand{\bbZ}{\mathbb{Z}}
\newcommand{\bbR}{\mathbb{R}}
\newcommand{\bbJ}{\mathbb{J}}
\newcommand{\bbC}{\mathbb{C}}
\newcommand{\bbG}{\mathbb{G}}
\newcommand{\bk}[1]{\langle #1 \rangle}
\newcommand{\sfo}{\mathsf{o}}
\newcommand{\sfO}{\mathsf{O}}
\newcommand{\icl}{\nu}
\newcommand{\iclb}{\icl_{\beta}}
\newcommand{\iclbh}{\icl_{\beta,h}}
\newcommand{\betac}{\beta_{\rm c}}
\newcommand{\betasat}{\beta_{\rm sat}}

\newcommand{\unitBall}{\mathscr{U}}

\begin{document}
	
	\title{Failure of Ornstein--Zernike asymptotics for the pair correlation function at high temperature and small density}
	
	\author{Yacine Aoun}
	\email[]{Yacine.Aoun@unige.ch}
	\affiliation{Section de mathématiques, Université de Genève, Switzerland}
	
	\author{Dmitry Ioffe}
	\affiliation{Faculty of IE\&M, Technion, Israel}
	
	\author{Sébastien Ott}
	\email[]{ott.sebast@gmail.com}
	\affiliation{Section de mathématiques, Université de Genève, Switzerland}
	
	\author{Yvan Velenik}
	\email[]{Yvan.Velenik@unige.ch}
	\affiliation{Section de mathématiques, Université de Genève, Switzerland}
	
	\date{\today}
	
	\begin{abstract}
		We report on recent results that show that the pair correlation function of systems with exponentially decaying interactions can fail to exhibit Ornstein--Zernike asymptotics at all sufficiently high temperatures and all sufficiently small densities.
		This turns out to be related to a lack of analyticity of the correlation length as a function of temperature and/or density and even occurs for one-dimensional systems.
	\end{abstract}
	
	\maketitle
	
In 1914, Ornstein and Zernike~\cite{Ornstein+Zernike-1914} introduced their celebrated equation for the density-density pair correlation function \(G(x,y)\):
\begin{equation}\label{eq:OZ-Equation}
	G(x,y) = C(x,y) + \rho\int C(x,z) G(z,y)\,\dd z,
\end{equation}
where \(C\) denotes the direct correlation function and \(\rho\) the density.
Using~\eqref{eq:OZ-Equation}, Zernike~\cite{Zernike-1916} determined the asymptotic behavior of the pair correlation function of systems with short-range interactions (meaning here interactions decaying faster than some exponentially decreasing function of the distance):
\begin{equation}\label{eq:OZ-Asymptotics}
	G(0,x) \sim B_d\frac{e^{-\nu \normII{x}}}{\normII{x}^{(d-1)/2}},
\end{equation}
as the Euclidean norm \(\normII{x}\) tends to infinity.
His derivation relied on a certain \emph{mass gap} assumption: \(|C(0,x)|\leq e^{-m\normII{x}} |G(0,x)|\), which was believed to hold in wide generality for short-range models at low density~\cite{Fisher-1964}.

It is usually expected that OZ behavior as displayed in~\eqref{eq:OZ-Asymptotics} should hold for general simple fluids with short-range interactions at sufficiently high temperature and/or sufficiently small density.
In this letter, we report on recent results that show that such is not necessarily the case. As will be explained below, this failure can be seen as resulting from a violation of the mass gap assumption.

For the sake of simplicity of exposition, we restrict our presentation to ferromagnetic Ising models on \(\bbZ^d\), but our results should hold in wide generality and have been established for other models in~\cite{Aoun+Ioffe+Ott+Velenik-2021} (including the self-avoiding walk, Potts models, Bernoulli percolation, the massive Gaussian Free Field and the XY model).
We start by discussing a simplified setting and then turn to a more general framework, in which some interesting new aspects emerge.
A heuristic explanation of the underlying mechanism, relating the latter to a suitable condensation phenomenon, is also provided.

\section{Overview in a simplified setting}

An anisotropic version of the OZ asymptotics~\eqref{eq:OZ-Asymptotics} has been rigorously derived for finite-range ferromagnetic Ising models on \(\bbZ^d\), both above the critical temperature~\cite{Campanino+Ioffe+Velenik-2003, Campanino+Ioffe+Velenik-2003b} and when the magnetic field is non-zero~\cite{Ott2018}: in both cases, as \(\normII{x}\) tends to infinity,
\begin{equation}\label{eq:OZ-Asymp-Rigorous}
	\bk{\sigma_0\sigma_x}_{\beta,h} - \bk{\sigma_{0}}_{\beta,h}\bk{\sigma_x}_{\beta,h} = \frac{A_{\beta,h}(\hat x)}{\normII{x}^{\frac{d-1}{2}}} e^{-\iclbh(\hat{x}) \normII{x}} (1+\sfo(1)),
\end{equation}
where both \(A_{\beta,h}\) and the inverse correlation length \(\iclbh\) are positive, analytic functions of the direction \(\hat{x} = x/\normII{x}\).
The same behavior still holds as long as the interaction decays super-exponentially~\cite{Aoun+Ott+Velenik-2021}.
Of course, this cannot be true for interactions decaying sub-exponentially, since the pair correlation function cannot decay slower than the interaction, as seems to have been first noted by Widom~\cite{Widom-1964}.

Let us thus restrict our attention to coupling constants \(J_{x,y} = J_{y-x}\) of the form
\begin{equation}\label{eq:Special-J}
	J_{x} = \normII{x}^\alpha e^{-c\normII{x}},
\end{equation}
for some \(\alpha\in\bbR\) and \(c>0\).
(More general forms will be considered below.)
Let us also assume for simplicity that there is no magnetic field (and omit it from the notation).
We emphasize, however, that the analysis below can also be done by fixing some arbitrary \(\beta\) and looking at the behavior of \(h\mapsto\iclbh(s)\).

Let \(s\) be a unit vector in \(\bbR^d\) and denote the inverse correlation length in direction \(s\) by
\begin{equation*}
	\iclb(s) = - \lim_{n\to\infty} \frac1n \ln \bk{\sigma_0\sigma_{[ns]}}_{\beta},
\end{equation*}
where \([ns]\) is a vertex of \(\bbZ^d\) which is closest to \(ns\in\bbR^d\).
It is known~\cite{Aizenman+Barsky+Fernandez-1987} that \(\iclb(s)>0\) for all \(\beta < \betac\), where \(\betac\) denotes the inverse critical temperature.
In addition, it is easy to prove that \(\iclb(s)\) is decreasing in \(\beta\) and satisfies
\begin{equation}\label{eq:ICL-InfiniteT}
	\lim_{\beta\to 0} \iclb(s) = c\normII{s} = c.
\end{equation}
Moreover, \(\lim_{\beta\uparrow\betac}\iclb(s)=0\)~\cite{McBryan+Rosen-1976}.
The behavior in~\eqref{eq:ICL-InfiniteT} is in sharp contrast with what happens when the interaction decays super-exponentially (in which case \(\lim_{\beta\to 0} \iclb(s) = \infty\)) and opens up the possibility that \(\iclb(s)\) \emph{saturates} at a finite temperature: \(\iclb(s) = c\normII{s} = c\) for all \(\beta<\betasat(s)\), for some \(\betasat(s) \in (0, \betac)\).

Observe that, if the latter were to occur, then there would be several remarkable consequences:
\begin{itemize}
	\item The inverse correlation length \(\iclb(s)\) would not depend analytically on the temperature in the whole high-temperature regime \((0, \betac)\).
	\item One of the basic assumptions of OZ theory, namely that the rate of decay of the pair correlation function is strictly slower than the rate of decay of the interaction (in the present case, \(\iclb(s) < c\normII{s}\)) would be violated.
\end{itemize}

It turns out that \emph{such a saturation phenomenon can occur, in any dimension}.
Whether or not it does so depends on the value of the exponent \(\alpha\).
Namely, saturation occurs (that is, \(\betasat(s)>0\)) if and only if
\begin{equation}\label{eq:Condition-Alpha}
	\alpha < -(d+1)/2 .
\end{equation}
In particular, even when \(d=1\), the inverse correlation length is a non-analytic function of \(\beta\) whenever \(\alpha < -1\), which contrasts with the well-known analyticity of all thermodynamic quantities~\cite{Ruelle-1975, Dobrushin-1974}.

As mentioned above, in the whole saturation regime \(\beta<\betasat(s)\), one of the basic assumptions of the OZ theory fails to hold.
In fact, one can show~\cite{Aoun+Ott+Velenik-2021} that, while the OZ asymptotic behavior~\eqref{eq:OZ-Asymp-Rigorous} still holds whenever \(\beta\in (\betasat(s), \betac)\), it fails when \(\beta<\betasat(s)\), where the pair correlation decays proportionally to the coupling constant.

\section{General setting}
Let us now consider coupling constants of the form
\begin{equation}\label{eq:GeneralJ}
	J_x = \psi(x) e^{-\normInt{x}},
\end{equation}
where \(\normInt{\cdot}\) is an arbitrary norm on \(\bbR^d\) and the prefactor \(\psi\) is positive and sub-exponential, in the sense that \(\lim_{\normInt{x}\to\infty} \ln(\psi(x))/\normInt{x} = 0\). We also assume that it behaves in an approximately isotropic manner: there exists \(c_{1}>0\) and a function \(\psi_0:\bbR_{>0}\to\bbR_{>0}\) such that \(\frac1c_{1}\psi_0(\normII{x}) \leq \psi(x) \leq c_{1}\psi_0(\normII{x})\) for all \(x\neq 0\).

\begin{figure}
	\includegraphics[width=6cm]{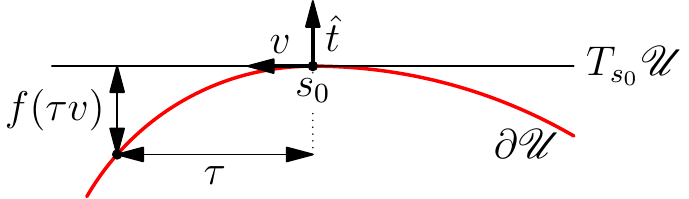}
	\caption{Local parametrization of \(\partial\unitBall\).}
	\label{fig:Parametrization}
\end{figure}

In order to ease the formulation of the extension of~\eqref{eq:Condition-Alpha} to this more general framework, we need to make an additional genericity assumption on the direction \(s\).
Let \(\unitBall\) be the unit-ball associated to the norm \(\normInt{\cdot}\). Given a direction \(s\), consider the corresponding point \(s_0 = s/\normInt{s}\in\partial\unitBall\).
By convexity, every \(u\in\partial\unitBall\) close enough to \(s_{0}\) can be written as 
\begin{equation}\label{eq:Parametrization}
u=s_{0}+\tau v-f(\tau v)\hat{t},
\end{equation}
where \(\tau>0\) is some small real number, \(v\) is unit vector in the tangent plane to \(\partial\unitBall\) at \(s_{0}\), \(\hat{t}\) is a unit vector normal to the tangent plane and \(f\) is some non-negative convex function defined on the tangent plane such that \(f(0)=0\) (see Figure~\ref{fig:Parametrization}).

Our assumption on the direction \(s\) is that the qualitative behavior of \(f\) is the same in all directions \(v\): there exist \(c_{2}>0\) and a non-decreasing convex function \(g\) such that, for any unit vector \(v\) in the tangent plane and \(\tau\) small enough, we have
\begin{equation}\label{eq:QuasiIsotropy}
	\dfrac{1}{c_{2}}g(\tau)\leq f(\tau v)\leq c_{2}g(\tau).
\end{equation} 
When this holds, we will say that \(\partial\mathsf{U}\) is quasi-isotropic in direction \(s\). Examples when quasi-isotropy is satisfied in every direction \(s\) include \(p\)-norms with \(p\in[1,+\infty]\).

Our main result on whether \(\betasat(s) > 0\) is the following: 
\newtheorem{theorem}{Theorem}
\begin{theorem}\label{thm:SaturationCondition}
Fix a direction \(s\)	in which \(\partial\mathsf{U}\) is quasi-isotropic.
Then, saturation occurs in direction \(s\) (that is, \(\betasat(s)>0\)) if and only if 
\begin{equation}\label{eq:Condition-General}
	\sum_{\ell\geq 1} \psi_0(\ell) (\ell g^{-1}(1/\ell))^{d-1} < \infty .
\end{equation}
\end{theorem}
(In~\eqref{eq:Condition-General}, we make the convention \(g^{-1}\equiv 1\) when \(g\equiv 0\).)

When considering~\eqref{eq:Special-J}, we can take \(\psi_0(x) = \normII{x}^\alpha\) and \(g(\tau) = \tau^2\); it is thus clear that~\eqref{eq:Condition-General} reduces to~\eqref{eq:Condition-Alpha} in that case.

\section{An example}
In this section, we illustrate on a specific example the direction-dependence of the saturation phenomenon.

Let us consider the model on \(\bbZ^2\) with interaction
\begin{equation*}
	J_x = \normp{x}^\alpha e^{-\normp{x}},
\end{equation*}
where \(\normp{\cdot}\) is the \(p\)-norm and we assume that \(p\in (2,\infty)\).

Let \(s\) be a unit vector in \(\bbR^2\) and let us write \(s_0 = (x_0, y_0) = s/\normp{s}\).
When both \(x_0\) and \(y_0\) are nonzero, the local parametrization~\eqref{eq:Parametrization} of \(\partial\unitBall\) at \(s_0\) applies with
\begin{equation*}
	f(\tau v) = \frac{p-1}2 \frac{x_0^{p-2}y_0^{p-2}}{(x_0^{2p-2} + y_0^{2p-2})^{3/2}} \tau^2 + \sfo(\tau^2),
\end{equation*}
so that one can choose \(g(\tau) = \tau^2\). Condition~\eqref{eq:Condition-General} then implies that there is saturation in direction \(s\) if and only if \(\alpha < -3/2\), as in~\eqref{eq:Condition-Alpha} (which should not be surprising, since \(\partial\unitBall\) has the same qualitative behavior in both cases).

In the remaining cases (\(s=\pm(1, 0)\) or \(\pm(0, 1)\)), the local parametrization~\eqref{eq:Parametrization} of \(\partial\unitBall\) at \(s_0\) applies with
\begin{equation*}
	f(\tau v) = \tfrac1p \tau^p + \sfo(\tau^p),
\end{equation*}
so that one can choose \(g(\tau) = \tau^p\). Therefore, in this case, there is saturation in direction \(s\) if and only if
\begin{equation*}
	\alpha < \frac{1}{p} - 2. 
\end{equation*}
This clearly shows that the occurrence of saturation depends on both the temperature and the direction. In particular, when \(-3/2 > \alpha \geq \frac{1}{p} - 2\), \(\betasat(s)=0\) when \(s\in \{\pm (1, 0), \pm (0, 1)\}\) but is positive for all other directions.

\section{An underlying condensation phenomenon}

\begin{figure}
\centering
	\includegraphics[width=3cm,height=4.2cm]{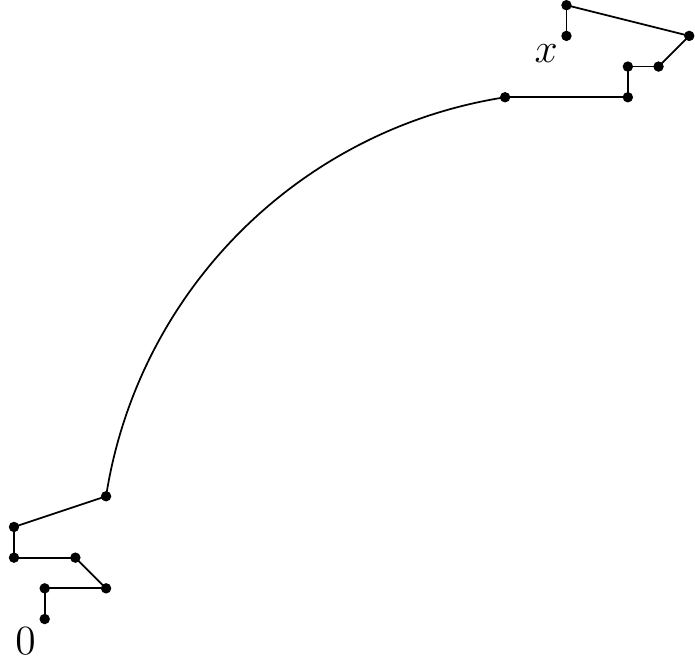}
	\hspace{1cm}
	\includegraphics[width=3cm,height=4.2cm]{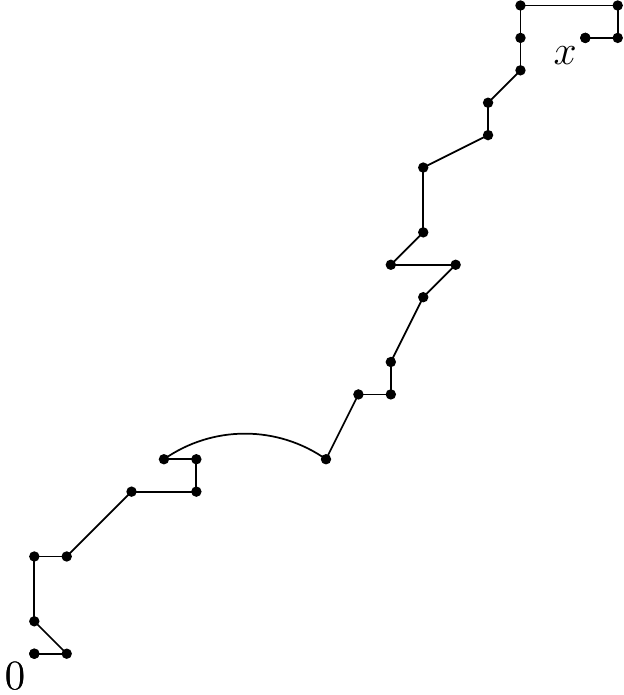}
	\caption{Sketches of a typical path \(\gamma\) in both regimes. Left: $\gamma$ consists of one giant step for $\beta<\betasat(\hat{x})$. Right: $\gamma$ consists of \(\sfO(\normII{x})\) small steps for $\beta>\betasat(\hat{x})$.}
	\label{fig:condensation}
\end{figure}

When \(\beta<\betac\), the 2-point function of the Ising model admits a graphical representation as a sum over paths:
\begin{equation}\label{eq:SumOverPaths}
	\bk{\sigma_0\sigma_x}_\beta = \sum_{\gamma:\, 0\to x} \mathsf{q}_\beta(\gamma),
\end{equation}
where the sum is over edge-self-avoiding paths connecting \(0\) to \(x\) and $\mathsf{q}_\beta(\gamma)$ is a suitable non-negative weight~\cite{Pfister+Velenik-1999}.
Suppose that \(\betasat(\hat x)>0\).
It turns out that the saturation phenomenon occurring at \(\betasat(\hat x)\) corresponds to a change of behavior of the typical paths contributing to the sum in~\eqref{eq:SumOverPaths}.
When \(\beta \in (\betasat(\hat x), \betac)\), typical paths \(\gamma\) have a length of order \(\normII{x}\) and behave diffusively, whereas
when \(\beta \in (0, \betasat(\hat x))\), typical paths \(\gamma\) contain \(\sfO(1)\) steps: there is a giant step connecting a vertex close to \(0\) to a vertex close to \(x\), see Fig.~\ref{fig:condensation}.
This change of behavior is reminiscent of condensation phenomena for sums of random variables; we refer to~\cite{Godreche-2019} for a review.

\section{Proof in dimension 1} 
	
In this section, we prove Theorem~\ref{thm:SaturationCondition} when \(d=1\). The proof of the full Theorem~\ref{thm:SaturationCondition} can be found in~\cite{Aoun+Ioffe+Ott+Velenik-2021}.

We consider the Ising model on \(\bbZ\) with interactions as in~\eqref{eq:GeneralJ}; for simplicity, we assume that \(\psi(y)=\psi(|y|)\) for all \(y\) and that \(\psi\) is monotone in \(\bbR_{>0}\). Introduce the generating functions (Laplace transforms)
\begin{equation*}
	\bbJ(\lambda) = \sum_{x} J_{x}e^{\lambda x},\quad \bbG(\lambda) = \sum_{x} \bk{\sigma_0\sigma_x}_\beta e^{\lambda x}.
\end{equation*}
Note that \(\bbG\) has radius of convergence \(\nu_{\beta}(1)\), while \(\bbJ\) has radius of convergence \(|1|\). Moreover, since we are assuming that \(d=1\), condition~\eqref{eq:Condition-General} is equivalent to \(\sum_{y\in\bbZ} \psi(y) < \infty\) and to \(\bbJ(|1|)<\infty\).

Our analysis below will be based on the following inequalities satisfied by the weights in~\eqref{eq:SumOverPaths} (see~\cite{Pfister+Velenik-1999}): there exists \(C_{\beta}>0\) such that, if \(\gamma=(\gamma_0, \gamma_1, \dots, \gamma_n)\),
\begin{equation}\label{eq:bound_on_q}
\beta^n \prod_{i=1}^n J_{\gamma_i-\gamma_{i-1}}
\geq
\mathsf{q}_\beta(\gamma)
\geq
\prod_{i=1}^n C_{\beta} J_{\gamma_i-\gamma_{i-1}}.
\end{equation}
Note that the lower bound immediately implies that \(\nu_{\beta}(\hat{x})\leq\normInt{\hat{x}}\) (simply consider the path \(\gamma\) composed of a single step from \(0\) to \(x\) in~\eqref{eq:SumOverPaths}).

\subsection{Proof that~\eqref{eq:Condition-General} implies \(\betasat>0\)}
Using~\eqref{eq:SumOverPaths} and the upper bound~\eqref{eq:bound_on_q}, one straightforwardly obtains
\begin{equation*}
	\bbG(\lambda)\leq  \frac{\beta\bbJ(\lambda)}{1-\beta \bbJ(\lambda)},
\end{equation*}
whenever \(\beta \bbJ(\lambda)<1\). So that if \(\bbJ(\lambda)<\infty\) and \(\beta < \bbJ(\lambda)^{-1}\), \(\lambda\) is in the convergence domain of \(\bbG\). As condition~\eqref{eq:Condition-General} implies that \(\bbJ(|1|)<\infty\), it follows that \(\nu_{\beta}(1)\geq |1|\) for \(\beta\) small enough.

\subsection{A comment on OZ equation at low density}
Let us pause a moment to make a comment about the previous argument. Consider the OZ equation~\eqref{eq:OZ-Equation}. Taking \(\bbG\) and \(\bbC\) the Laplace transforms of \(x\mapsto G(0,x)\) and \(x \mapsto C(0,x)\), one obtains the relation \(\bbG  = \frac{\bbC}{1-\rho \bbC}\). For \(d=1\), \(C(0,x)\geq 0\) (ferromagnetic assumption), and supposing that the asymptotic decay of \(C\) coincides with the one of \(J\) (which is for example the case in the Percus–Yevick approximation), the radii of convergence of \(\bbG,\bbC\) are \(\nu(1)\) and \(|1|\) respectively. Moreover, the same argument as previously implies that when \(\bbC(|1|)<\infty\) and \(\rho\) is small enough, \(\bbG(|1|)<\infty\) and therefore that \(|1|\leq\nu(1) \) (failure of mass gap). In higher dimensions, the same argument (with more involved considerations about the link between the convergence domains of \(\bbG,\bbC\) and \(\nu,|\ |\)) is at the heart of the general condition~\eqref{eq:Condition-General}.

\subsection{Proof that \(\sum_y \psi(y) = +\infty\) implies \(\betasat=0\)}
Since \(\psi\) is non-summable, there exists \(R\) such that
\begin{equation*}
\sum_{y=1}^R \psi(y) \geq \frac{e}{C_\beta}.
\end{equation*}
Let \(x\in\mathbb{N}\) and \(N = \normII{x}/R\).
We have
\begin{align*}
\bk{\sigma_0\sigma_x}_\beta 
&\geq 
\sum_{y_1=1}^R \cdots \sum_{y_N=1}^R \Bigl( \prod_{i=1}^{N} C_{\beta} J_{y_i} \Bigr) J_{x-(y_1+\dots+y_N)}\\
&\geq
e^{-\normInt{x}} \sum_{y_1=1}^R \cdots \sum_{y_N=1}^R \Bigl( \prod_{i=1}^{n} C_\beta \psi(y_i) \Bigr) \psi\bigl(x-\sum_i y_i\bigr) \\
&=
e^{-\normInt{x}-\sfo(\normII{x})} \Bigl( C_\beta \sum_{y=1}^R \psi(y_i) \Bigr)^{\! N} \\
&\geq
e^{-\normInt{x}-\sfo(\normII{x})} e^{N}
=
e^{-\normInt{x} + \frac{\normII{x}}{R}-\sfo(\normII{x})},
\end{align*}
where we used \(\normInt{x} = \normInt{y_1+\dots+y_N}\) in the second line; the fact that \(\psi\) is sub-exponential in the third line; the choice of \(R\) in the fourth line.
This immediately implies that \(\iclb(\hat x) \leq \normInt{\hat x} - \frac1R\).
Since \(\beta \in (0,\betac)\) was arbitrary, we conclude that \(\betasat=0\).

\section{Failure of Ornstein-Zernike asymptotics}

In this section, we illustrate, in a simple example, the mechanism leading to a violation of OZ asymptotics when saturation occurs. Let us thus consider the Ising model on $\mathbb{Z}$ with coupling constants of the form
\begin{equation*}
	J_{y} = \normInt{y}^\alpha e^{-\normInt{y}}.
\end{equation*}
(Again, we refer to~\cite{Aoun+Ioffe+Ott+Velenik-2021} for a treatment of the general case in any dimension.)

In the previous section, it was shown that saturation occurs when
\begin{equation}\label{eq:summability_polynomial}
	\sum_{y\in\mathbb{Z}\setminus\lbrace 0\rbrace}\normInt{y}^\alpha <\infty.
\end{equation}
We are going to show that there exists \(c_{\beta}>0\) such that
\begin{equation}\label{eq:failure_OZ}
	\dfrac{1}{c_{\beta}} \normInt{x}^\alpha e^{-\normInt{x}}
	\leq
	\bk{\sigma_0\sigma_x}_\beta
	\leq 
	c_{\beta} \normInt{x}^\alpha e^{-\normInt{x}}.
\end{equation}
The lower bound follows directly from~\eqref{eq:SumOverPaths} by considering the path composed of a single step from $0$ to $x$. For the upper bound, note first that~\eqref{eq:bound_on_q} implies 
\begin{align}
	\bk{\sigma_0\sigma_x}_\beta 
	&\leq 
	\sum_{n=1}^{\infty}\sum_{\substack{y_1, \dots, y_n\in\bbZ\setminus\{0\} \\ y_1+\dots+y_n=x}} 
	\beta^n \prod_{i=1}^{n} \normInt{y_i}^\alpha e^{-\normInt{y_i}} \notag \\
	&\leq 
	e^{-\normInt{x}}
	\sum_{n=1}^{\infty} n \sum_{\substack{y_1, \dots, y_n\in\bbZ\setminus\{0\} \\ y_1+\dots+y_n=x \\ \normInt{y_1}\geq\max_{i}\normInt{y_i}}} \beta^n \prod_{i=1}^{n} \normInt{y_i}^{\alpha}, \label{eq:BoundOn2ptfFailureOZ}
\end{align}
where in the last line, we assumed that \(y_1\) is the vector with maximal \(\normInt{\cdot}\)-norm and used that \(\normInt{y_1}+\dots+\normInt{y_n} \geq \normInt{y_1+\dots+y_n} = \normInt{x}\).  Observe now that, for any two real numbers \(a, b\) with \(a\geq b\), we have 
\begin{equation*}
	a^{\alpha} b^{\alpha}
	\leq \biggl(\frac{a+b}{2}\biggr)^{\!\!\alpha} b^\alpha
	= 2^{-\alpha} (a+b)^{\alpha} b^{\alpha},
\end{equation*}
since \(\alpha\) is negative (by~\eqref{eq:summability_polynomial}).
Using this observation \(n-1\) times, it follows that 
\begin{align}
	\normInt{y_1}^{\alpha} \prod_{i=2}^{n} \normInt{y_i}^{\alpha}
	&\leq 
	\Bigl(\sum_{i=1}^{n}\normInt{y_i}\Bigr)^{\! \alpha} \prod_{i=2}^{n} 2^{-\alpha} \normInt{y_i}^{\alpha} \notag\\
	&\leq 
	\normInt{x}^{\alpha} \prod_{i=2}^{n} 2^{-\alpha} \normInt{y_i}^{\alpha}, \label{eq:SimpleBound}
\end{align}
where we used \(\normInt{y_1}\geq \max_i\normInt{y_i}\) for the first inequality and \(\normInt{y_1}+\dots+\normInt{y_n} \geq \normInt{x}\) and \(\alpha<0\) for the second one.

Using~\eqref{eq:SimpleBound} in~\eqref{eq:BoundOn2ptfFailureOZ} yields
\begin{align*}
	\bk{\sigma_0\sigma_x}_\beta 
	&\leq 
	\normInt{x}^{\alpha} e^{-\normInt{x}} \sum_{n=1}^{\infty} n \!\!\!\! \sum_{\substack{y_1, \dots, y_n\in\bbZ\setminus\{0\} \\ y_1+\dots+y_n=x \\ \normInt{y_1}\geq\max_{i}\normInt{y_i}}}\!\!\!\!
	(2^{-\alpha}\beta)^{n} \prod_{i=2}^{n} \normInt{y_i}^{\alpha} \\
	&\leq 
	\normInt{x}^{\alpha} e^{-\normInt{x}} 2^{-\alpha}\beta \sum_{n=1}^{\infty} n \Bigl(2^{-\alpha}\beta\sum_{y\in\mathbb{Z}\setminus\lbrace 0\rbrace} \normInt{y}^{\alpha}\Bigr)^{\! n-1}.
\end{align*}
Thanks to Condition~\eqref{eq:summability_polynomial}, we conclude that the sum over \(n\) is convergent when \(\beta\) is small enough, which yields the upper bound in~\eqref{eq:failure_OZ}.

~\section{Conclusion}

We have considered an Ising model with exponentially decaying interactions and have explained how the latter can, in suitable circumstances, undergo a saturation ``transition'', in which the rate of exponential decay of the pair correlation function coincides with the rate of exponential decay of the interaction over a non-empty interval of temperatures \((0,\betasat)\) (where \(\betasat\in (0,\betac)\) is in general direction-dependent). Since the inverse correlation length becomes independent of the temperature below \(\betasat\), it is not an analytic function of \(\beta\) on \((0, \betac)\). 
When this occurs, the pair correlation function fails to exhibit Ornstein--Zernike asymptotics for all \(\beta<\betasat\).

A number of issues remain open.
We have limited our analysis, for technical reasons, to ferromagnetic systems.
It is unclear to us whether the general phenomenology changes when this assumption is dropped.
In addition, the behavior of the system at $\betasat$ is not understood, although it can be shown that in one-dimensional ferromagnetic systems, saturation does not occur at $\betasat$~\cite{Aoun+Ott+Velenik-2021}.

\acknowledgments
YA thanks Hugo Duminil-Copin for financial support. YV was partially supported by the Swiss National Science Foundation through the NCCR SwissMAP.

\bibliography{AIOV21-PRE}

\end{document}